\documentclass{Interspeech2024}

 \usepackage{multirow,pifont} 
\usepackage{subcaption, tablefootnote,graphicx}  

\interspeechcameraready

\title{Streaming Decoder-Only Automatic Speech Recognition with Discrete Speech Units: A Pilot Study}

\name[affiliation={1,2}]{Peikun}{Chen}
\name[affiliation={2}]{Sining}{Sun}
\name[affiliation={2}]{Changhao}{Shan}
\name[affiliation={2}]{Qing}{Yang}
\name[affiliation={1*}]{Lei}{Xie}

\address{
  $^1$Audio, Speech and Language Processing Group (ASLP@NPU), \\ Northwestern Polytechnical University, Xi'an, China\\
  $^2$Du Xiaoman Financial, China}
\email{cpk00925@mail.nwpu.edu.cn, lxie@nwpu.edu.cn\thanks{$^*$: Corresponding author.}}

\keywords{streaming automatic speech recognition, discrete-token, decoder-only Transformer}

\begin{document}

\maketitle

\begin{abstract}
Unified speech-text models like SpeechGPT, VioLA, and AudioPaLM have shown impressive performance across various speech-related tasks, especially in Automatic Speech Recognition (ASR). 
These models typically adopt a unified method to model discrete speech and text tokens, followed by training a decoder-only transformer. 
However, they are all designed for non-streaming ASR tasks, where the entire speech utterance is needed during decoding.
Hence, we introduce a decoder-only model exclusively designed for streaming recognition, incorporating a dedicated boundary token to facilitate streaming recognition and employing causal attention masking during the training phase.
Furthermore, we introduce right-chunk attention and various data augmentation techniques to improve the model's contextual modeling abilities.
While achieving streaming speech recognition, experiments on the AISHELL-1 and -2 datasets demonstrate the competitive performance of our streaming approach with non-streaming decoder-only counterparts. The code we used for this work can be found here\footnote{\url{https://github.com/chenpk00/IS2024_stream_decoder_only_asr}}.
\end{abstract}

\section{Introduction}


Recently, large language models (LLMs)~\cite{gpt, llama, glm, qwen} have made great progress in various natural language processing (NLP) tasks. Methods based on LLMs are also leading a revolution in other research fields, such as computing vision~\cite{videogpt, cv1, cv2} and speech~\cite{viola, speechgpt, audiopalm, laura}. Thanks to the powerful understanding of language and generalization ability of an LLM, transferring LLM pretrained from massive amounts of text data to speech recognition tasks can also bring significant word error rate reduction~\cite{DBLP:journals/corr/abs-2401-10446, DBLP:journals/corr/abs-2402-05457}. As for the model architecture, most of the LLMs are based on decoder-only transformers~\cite{transformer}, which simplifies the model structure used for ASR tasks. Therefore, LLM-based ASR is becoming a hot research topic in the field. 

Currently, there are two primary approaches to integrating the speech modality with LLMs in LLM-based ASR.
One approach involves directly integrating continuous features with text embedding through a trainable adaptor~\cite{qwenaudio, salm, wu}. This kind of approach introduces additional acoustic encoders and models speech and text separately.
Conversely, the other approach is to treat speech representation as textual tokens and employ a decoder-only model to optimize multi-modal tasks effectively.
For instance, VioLA~\cite{viola} converts continuous speech signals to discrete codec codes via EnCodec~\cite{encodec} and unifies several speech-related tasks into a conditional language modeling task. SpeechGPT~\cite{speechgpt} employs LLaMA~\cite{llama} as its foundational framework, utilizing $k$-means clustering derived from Hubert~\cite{hubert} to tokenize speech signal. Similarly, AudioPaLM~\cite{audiopalm} utilizes PaLM-2~\cite{palm} as its underlying architecture and extracts discrete tokens from the encoder of Universal Speech Model~\cite{usm}.
Such a unified modeling approach has been demonstrated to effectively improve the ASR performance. The decoder-only transformer model with unified discrete input provides a new paradigm for various speech-related tasks, including but not limited to speech recognition and speech synthesis.


Previous works on decoder-only ASR tasks mainly focused on non-streaming scenarios ~\cite {viola, audiopalm}.
However, real-time streaming recognition can give faster recognition results and a better user experience in real-world applications. Many works have been proposed to build a faster and better streaming ASR system in the past decades based on various end-to-end speech recognition frameworks~\cite{semi,stream2}.  However, the exploration of streaming decoder-only speech recognition is very limited. 
As decoder-only-based ASR model performance improves and the number of model parameters increases, streaming low-latency inference becomes a challenging task.

This paper presents a pilot study on the streaming decoder-only transformer ASR model. Current non-streaming decoder-only transformer ASR models learn to predict text tokens autoregressively using the whole speech utterance~\cite{lossmask}. For the streaming version, it is necessary to emit text tokens with minimal delay as the corresponding speech segment is received.
To this end, we investigate two approaches based on the speech-to-text alignment obtained by a GMM-HMM model~\cite{gmm}. Specifically, the first approach, \textbf{T}ext \textbf{T}oken \textbf{I}nsertion (TTI), inserts the corresponding text tokens into the speech token sequences directly under the guide of speech-to-text alignment during training. By contrast, in \textbf{B}oundary \textbf{T}oken \textbf{I}nsertion (BTI), special ``boundary tokens" are inserted into the speech token sequences in the same way with text tokens added at the end, effectively decoupling the speech and text modalities. Upon triggering a boundary token, the corresponding text token can be generated through a one-step inference process autoregressively.
Meanwhile, we introduce right-chunk attention and various data augmentation techniques to improve the streaming model's contextual modeling ability.
We also explore the efficacy of leveraging an off-the-shelf text LLM to initialize our streaming ASR model.

\begin{figure*}[htbp]  
    \centering  
    \begin{subfigure}[t]{0.29\textwidth}  
        \centering  
        \includegraphics[width=\textwidth]{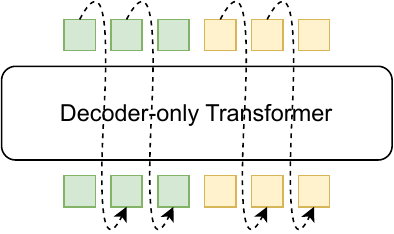}  
        \caption{Non-streaming model}  
    \end{subfigure}  
    \hfill
    \begin{subfigure}[t]{0.29\textwidth}  
        \centering  
        \includegraphics[width=\textwidth]{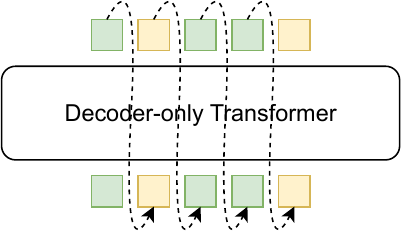}  
        \caption{TTI streaming model}  
    \end{subfigure}  
    \hfill
    \begin{subfigure}[t]{0.29\textwidth}  
        \centering  
        \includegraphics[width=\textwidth]{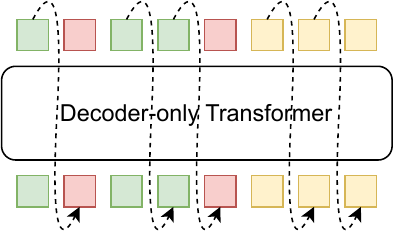}  
        \caption{BTI streaming model}  
    \end{subfigure}  
    \hfill
    \begin{subfigure}[t]{0.08\textwidth}  
        \centering  
        \includegraphics[width=\textwidth]{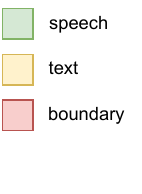} 
    \end{subfigure}  
    \caption{The comparison of different methods for training discrete-token-based decoder-only Transformer for ASR. (a) non-streaming model: decoding after receiving the whole speech token; (b) Text token insertion (TTI) streaming model: inserts text tokens into speech token sequences directly under the guide of speech-to-text alignment; (3) Boundary token insertion (BTI) streaming model: insert ``boundary tokens" into the discrete speech token sequence.}  
    \label{fig1}  
\end{figure*}

Experiments show that our proposed streaming decoder-only model can obtain 5.9\% and 7.2\% character error rate (CER) on two Chinese Mandarin corpora,  AISHELL-1~\cite{aishell1} and AISHELL-2~\cite{aishell2}, respectively, showing competitive performance with the non-streaming counterpart. Our results also show that the streaming decoder-only Transformer ASR model can benefit from the initialization from an off-the-shelf text LLM, such as Qwen~\cite{qwen}.

\section{Proposed Method}
\subsection{Streaming decoder-only model architecture}
Figure~\ref{fig1} illustrates three types of decoder-only models designed for ASR tasks based on discrete speech token input. Among them, Figure~\ref{fig1} (a) represents a non-streaming model proposed in~\cite{lossmask}, whereas Figures~\ref{fig1} (b) and (c) depict two variations of streaming frameworks. Both models (b) and (c) necessitate force alignment between speech and text.
As shown in equation (\ref{eq1}), given discrete speech token sequence $x=(x_1, ...,x_t, ...x_T)$ and the correspoing text token sequence $y=(y1,...,y_L)$, streaming ASR model is optimized by maximizing the conditional probability, where $t_{y_i+\Delta}$ is the time of emitting text token $y_i$, $\Delta$ is a constant, which means how many right context tokens can be used, $x_{\leq t_{y_i}+\Delta}=({x_1,...,x_{t_{y_i}+\Delta}})$ and $\theta$ is the trainable model parameter. 

\begin{equation}\label{eq1}
p(y|x; \theta) =  {\textstyle \prod_{1}^{L}} p\left(y_{i} \mid x_{\leq t_{y_i+\Delta}}, \theta\right).
\end{equation}

In Figure~\ref{fig1} (b), the Text Token Insertion (\textbf{TTI}) approach showcases the interleaving of discrete speech and text tokens, with text tokens inserted into speech tokens at the end of the corresponding speech segment. Mathematically, it equals to optimize equation (\ref{eq1}) directly. However, the mix of text and speech tokens complicates the use of beam search during decoding. During inference, speech tokens can be treated as conditions. Given the interleaved nature of text and speech tokens, triggering a text token during beam search necessitates caching all historical hidden states (e.g., key and value of self-attention) for each search path. 

Figure~\ref{fig1} (c) illustrates the Boundary Token Insertion (\textbf{BTI}) approach, where a special token, instead of a text token, is inserted into the speech token sequence, effectively decoupling the text and speech token sequences. In contrast to Figure~\ref{fig1} (b), this process can be viewed as comprising two stages: the first stage involves determining the boundary position, while the second stage entails predicting the corresponding specific text token conditioned on the history of speech tokens.  Equation (\ref{eq2}) provides a formal definition. Here, a hidden variable $b$ is introduced, where $b=(b_1,...,b_T) \in {0,1}^T$ represents one of the possible boundary paths, and $\beta$ denotes the set of all possible paths. However, in practice, optimizing by summing over all possible paths is computationally challenging. Therefore, we opt to approximate this optimization problem by selecting the most probable path $b_p$.

\begin{equation}
\begin{split}\label{eq2}
    p\left(y \mid x, \theta\right)
    &={\sum_{b\in \beta }}p\left(y,b \mid x, \theta\right) \\
    &={\sum_{b\in \beta }}{\textstyle \prod_{1}^{L}} p\left(y_{i},b \mid x\leq t_{y_i+\Delta}, \theta\right)  \\
    &\approx {\textstyle \prod_{1}^{L}} p\left(y_i \mid b_p,x_\leq t_{y_i+\Delta},\theta\right) p\left(b_p \mid x_\leq t_{y_i+\Delta}, \theta\right),  \\
\end{split}    
\end{equation}

\begin{figure}[htbp]
    \centering  
    \begin{subfigure}[t]{0.4\textwidth}  
        \centering  
        \includegraphics[width=\textwidth]{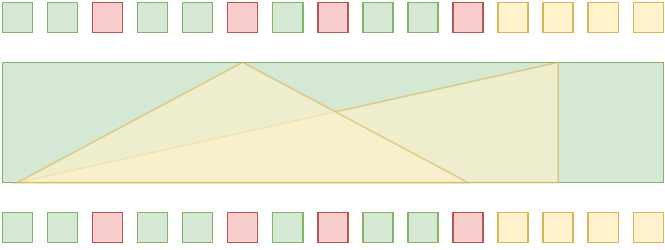}  
        \caption{global attention}  
    \end{subfigure}
    \hfill
    \begin{subfigure}[t]{0.4\textwidth}  
        \centering  
        \includegraphics[width=\textwidth]{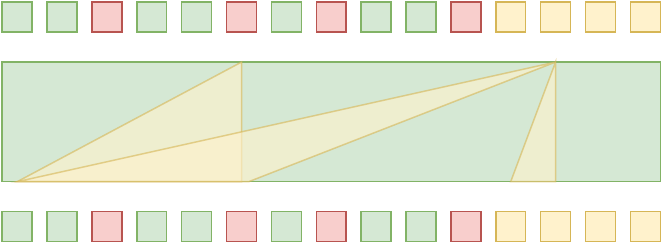}  
        \caption{causal attention}  
    \end{subfigure}
    \hfill
    \begin{subfigure}[t]{0.4\textwidth}  
        \centering  
        \includegraphics[width=\textwidth]{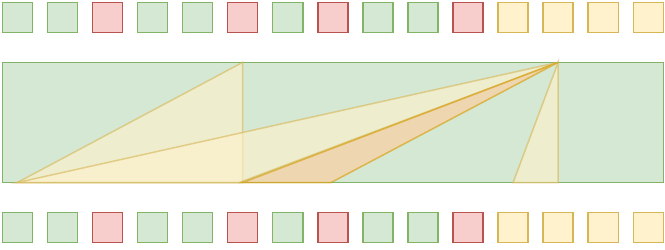}  
        \caption{right-chunk attention}  
    \end{subfigure} 
    \caption{Example diagram of different attention mechanisms. The green blocks indicate the part of the LLM. The yellow triangle indicates the part of the attention area. (a) global attention (b) causal attention; (c) right-chunk attention.}  
    \label{fig: att}  
\end{figure}

\subsection{Right-chunk attention}

The previous non-streaming decoder-only ASR models predict the text tokens aggressively with the entire discrete speech tokens. Illustrated in Figure \ref{fig: att} (a), the speech tokens primarily attend to each other, while the text tokens focus on all discrete speech tokens and preceding text tokens.

To achieve streaming speech recognition, we introduce a causal attention mechanism. As shown in the formula (\ref{eq3}), a casual mask is applied during self-attention calculation, where $M$ is the mask. 
\begin{equation}\label{eq3}
    \text{Attention}(Q, K, V) = \text{softmax}\left(\frac{QKM}{\sqrt{d_k}}\right){V},
\end{equation}
We divide attention mask $M$ into two parts,  denoted as $M_{speech}$ and $M_{text}$, representing the attention mask used for speech and text tokens, respectively. Equations (\ref{eq4}) and (\ref{eq5}) provide the definitions of $M_{\text{speech}}$ and $M_{\text{text}}$.
\begin{equation}\label{eq4}
    M_{speech}(i,j)=\left\{\begin{array}{ll}
    \text {True } & i \le j \\
    \text {False } & \text { otherwise }
    \end{array}\right.,
\end{equation}
where $0\le i\le T+L$ because of the insertion of boundary tokens in the speech.
Meanwhile, as shown in the figure~\ref{fig1} (b), $mask_{text}$ can be represented by
\begin{equation}\label{eq5}
\begin{split}
     M_{text}(i,j)=\left\{\begin{array}{ll}
    \text {True } & j \le t_{y_i} \\ &or \ T+L < i \le j \le T+2L\\
    \text {False } &\text {otherwise }
    \end{array}\right.,   
\end{split}
\end{equation}
where $M_{text}$ consists of two yellow triangles. At each time step,  the current text focuses on both the preceding speech and text.

Unlike non-streaming ASR models, streaming ASR models face limitations in considering global information, leading to weakened contextual modeling capabilities. To address this issue, we integrate the right-chunk attention mechanism, illustrated in Figure \ref{fig: att} (c). Unlike causal attention, right-chunk attention in the text portion enables capturing more speech information.
\begin{equation}
\begin{split}
     M_{text}(i,j)=\left\{\begin{array}{ll}
    \text {True } & j \le t_{y_i+\Delta} \\ &or \ T+L < i \le j \le T+2L\\
    \text {False } &\text {otherwise }
    \end{array}\right.,   
\end{split}
\end{equation}
where $\Delta$ means how many speech tokens in the right context can be used. Larger $\Delta$ also means a higher latency. 

\subsection{Data Pre-processing}
Another significant improvement lies in the pre-processing steps. Throughout the training and decoding stages, we observed a tendency for the model to more readily overfit discrete features compared to continuous features, thus significantly impacting the final results. To address this concern, we employed data pre-processing, which plays a crucial role in model training and can substantially enhance data diversity and quantity.
To improve the model's robustness, we implemented the following strategies:

\textbf{Speed Perturbation:} Adjusting the audio speed increases the variety of discrete speech tokens, allowing the text segment to match a wider range of token combinations.  In our study, we applied speed perturbation by altering the audio speed to 0.9 and 1.1 times the original speed.

\textbf{Trigger Shift:} Leveraging alignment biases to enhance the robustness of trigger tokens, we randomly shift trigger tokens by 1-4 frames with a probability of 30\% during the training phase.

\textbf{Time Masking:} Time masking is applied to input tokens other than trigger tokens, encompassing both speech and text tokens, by substituting each token with a special padding token with a probability of 0.3.

\textbf{Random De-duplication:} Employing a randomized de-duplication approach reduces computational complexity while amplifying data diversity. Concurrently, during the decoding phase, we implement global de-duplication to further alleviate computational overhead.

\textbf{Label Smoothing:} Discretized speech tokens manifest intersections within clustering, resulting in losses during the speech discretization process. To address this issue, we adopt label smoothing~\cite{ls}, 
\begin{equation}
\label{eq:ls}
\mathcal{L}_{\text{LS}} = \sum_{t=1}^{T_s} D_{\text{KL}}( q'(x_t|x_{<t}) || p(x_t|x_{<t}; \theta)  ) ,
\end{equation}
where $q'(x_t|x_{<t})$ is a soft label by label smoothing instead of a one-hot label, and $D_{\text{kl}}$ is the Kullback-Leibler divergence.

\section{Experiments}
\subsection{Experimental Setup}
Our implementation is based on Wenet~\cite{wenet}, an open-source toolkit for End-to-End (E2E) speech recognition. Our ASR models employ a decoder-only architecture based on the transformer. We have established two model configurations: a small model (70 million parameters) without Large Language Model (LLM) initialization, compared with the previous non-streaming AEDs model, and a large model (310 million parameters) used to verify the feasibility of off-the-shelf LLM initialization. In future endeavors, we intend to incorporate LLMs with significantly more parameters for validation, such as models with 2 billion or larger parameters.

\textbf{Dataset:}
In this study, we conduct experiments on two commonly used Chinese Mandarin corpora, 178-hour AISHELL-1~\cite{aishell1} and 1000-hour AISHELL-2~\cite{aishell2}.  We report the character error rates (CER) of various models.

\textbf{Discrete Speech Tokens:}
In this paper, we draw inspiration from a recent study~\cite{CCA} employing Canonical Correlation Analysis (CCA) to evaluate the similarity between layer representations and word labels. For the Chinese corpus, we select Chinese HuBERT large\footnote{\url{https://huggingface.co/TencentGameMate/chinese-hubert-large}\label{hu}}. Subsequently, we choose layer 21 from the large models, as it demonstrates the highest CCA similarities with word labels. The number of K-Means clusters is set to 2,000, consistent with the previous method~\cite{lossmask, chang}.

\textbf{Model Configuration:}
The small decoder-only transformer without LLM init comprises 8 blocks, each with 8 self-attention heads, with an attention dimension of 512, and a feed-forward network (FFN) with an intermediate hidden dimension of 1024.
To explore the effectiveness of using off-the-shelf LLM init, we also adopt Qwen\cite{qwen} as the backbone of our decoder-only Transformer model for discrete-token-based ASR systems. We use the Qwen2-0.5B\footnote{\url{https://huggingface.co/Qwen/Qwen1.5-0.5B}} model (transformer with 310M parameters), which consists of 24 layers with the hidden size 1024, 16 attention heads, max sequence length of 1024. 
Note that we do not use the Qwen text tokenizer because it can cause token sparsity problems, which means there is not enough training data for some tokens. Instead, we directly discretize text into Chinese characters. There are a total of 7000 model units with 5000 commonly used Chinese char and 2000 speech tokens.  In all experiments, we set the dropout rate to 0.1. Commonly during the training phase, we dynamically set $\Delta$ to be equal to the length of the speech segment corresponding to the next text token. During ASR decoding, we set the beam size to 10 and did not utilize language models in our experiments.

\begin{table}[]
  \caption{CERs (\%) on AISHELL-1 for different methods.}
  \label{tab: main result}
  \centering
\resizebox{0.46\textwidth}{!}{
\begin{tabular}{cccccc}
\hline
\multirow{2}{*}{ID} & \multirow{2}{*}{Feature} & \multirow{2}{*}{Model type} & \multirow{2}{*}{Streaming} &\multicolumn{2}{c}{CER} \\
        & &       &     &dev& test \\ \hline
B1& SSL\textsuperscript{~\ref{hu}}      & encoder-decoder &  \ding{55}           &3.8   &4.0    \\
B2&Fbank~\cite{ebran}    & encoder-decoder &  \ding{55}   &4.2   &4.5    \\
B3&Discrete~\cite{chang} & encoder-decoder &  \ding{55}   &4.6   &4.9    \\\hline
\multicolumn{6}{l}{\textit{Small}}            \\
S1&Discrete      & decoder-only    &  \ding{55}   &5.9   &6.2    \\ 
S2&Discrete (TTI) & decoder-only    &  \checkmark  &9.4   &9.8    \\
S3&Discrete (BTI) & decoder-only    &  \checkmark  &6.1   &6.4    \\ \hline
\multicolumn{6}{l}{\textit{Large (w/ Qwen-0.5B init)}}               \\
L1&Discrete      & decoder-only    &  \ding{55}   &5.2   &5.5    \\ 
L2&Discrete (TTI) & decoder-only    &  \checkmark  &9.2   &9.5    \\
L3&Discrete (BTI) & decoder-only    &  \checkmark  &5.6   &5.9    \\ \hline
\end{tabular}
}
\end{table}

\subsection{Main Results}

Table \ref{tab: main result} presents a CER on AISHELL-1. 
The first group in Table \ref{tab: main result} presents results from previous studies, including semi-supervised learning (SSL)~\cite{hubert}, E-Branchformer~\cite{ebran} based ASR model with continuous Fbank feature or discrete speech token~\cite{chang} as input. The second group displays the results of our small model trained with our proposed two streaming methods using random initialization. In comparison to S3, the substitution errors of S2 have notably increased due to TTI's design of text token insertion, which weakens its contextual modeling ability. It is evident that better model performance was achieved with S3 by decoupling boundary prediction and text prediction through BTI. The third group shows the results of the large model initializing with Qwen-0.5B LLM. With LLM initialization,  we first train a non-streaming decoder-only model, which is model L1. Then we fine-tune the model L1 with our proposed streaming methods TTI and BTI, resulting in two streaming versions, L2 and L3. We observe that the large model with LLM initialization achieves a 5.9\% CER on the AISHELL-1 test set. Upon comparing L2 with S2, it is noted that LLM initialization yields a relatively minor CER reduction for the TTI approach.  This is attributed to the differing nature of interleaving speech and text compared to LLMs, resulting in only a relative CER benefit of 3.1\% 
 (9.8\%$\to$9.5\%). Conversely, employing BTI results in a 9.2\% (6.5\%$\to$5.9\%) relative CER reduction on the test set compared to model S3. Notably, BTI not only exhibits superior CER performance but also demonstrates better adaptability to LLMs. Hence, in subsequent experiments, we employ the BTI approach.

\subsection{Ablation Study}

Table~\ref{tab: ablation} shows the impact of right-chunk attention and various data augmentation based on model S3. 
Notably, the right-chunk attention has the most significant impact on the overall CER. Without right-chunk attention, the CER increases from 6.5\% to 7.8\% on the AISHELL-1 test set. The absence of right-chunk attention results in more substitution errors due to the limited contextual information available.

Among the five data pre-processing methods, label smoothing is the most effective method, resulting in a relative 9.7\% CER reduction. We observe that when modeling speech and text tokens in a unified manner, predicting the next speech token is considerably easier than predicting the next text token. Consequently, the model tends to overfit during speech token prediction. Label smoothing effectively mitigates this overfitting issue. Additionally, other methods also contribute to some reduction in CER, ranging from 4.4\% to 7.2\%.

\begin{table}[]
  \caption{Ablation of each component’s impact on CERs (\%).}
  \label{tab: ablation}
  \centering
\begin{tabular}{lcc}
\hline
\multirow{2}{*}{Method}  &\multicolumn{2}{c}{CER} \\
                                &dev         & test \\ \hline
BTI (ours)                      &6.1         & 6.4     \\
\ \ w/o right-chunk attention   &7.4         & 7.8     \\
\ \ w/o speed perturb           &6.5         & 6.9     \\
\ \ w/o trigger shift           &6.4         & 6.7     \\
\ \ w/o time mask               &6.4         & 6.8     \\
\ \ w/o random de-duplication   &6.3         & 6.7     \\
\ \ w/o label smoothing         &6.8         & 7.2     \\ \hline
\end{tabular}
\end{table}

\subsection{Results on AISHELL-2}

In Table~\ref{tab: main result aishell2}, we present the performance on the AISHELL-2  corpus without speed perturbation. The top lines list the conventional FBank-based ASR system and the Hubert-Large discrete token-based ASR models. It is evident that using discrete token input with a decoder-only model yields slightly inferior performance compared to the encoder-decoder model (6.6\% vs 6.9\%). Meanwhile, the recognition accuracy of streaming models decreased by 4.1\% (6.9\%$\to$7.2\%) compared to non-streaming results. We find that training ASR models using the discrete units on large-scale data can be quite efficient. We believe that as the amount of data and model parameters increase, the decoder-only model can completely surpass the traditional encoder-decoder model. 

\begin{table}[]
  \caption{CERs (\%) on AISHELL-2 for different methods.}
  \label{tab: main result aishell2}
  \centering
\resizebox{0.46\textwidth}{!}{
\begin{tabular}{ccccc}
\hline
ID &Feature & Model type & Streaming &CER  \\ \hline
B4&Fbank\tablefootnote{\url{https://github.com/wenet-e2e/wenet}}    & encoder-decoder &  \ding{55}   &6.2       \\
B5&Discrete & encoder-decoder &  \ding{55}   &6.6       \\\hline
\multicolumn{4}{l}{\textit{Large (w/ Qwen-0.5B init)}}               \\
L4&Discrete & decoder-only    &  \ding{55}   &6.9       \\ 
L5&Discrete (BTI) & decoder-only    &  \checkmark  &7.2       \\ \hline
\end{tabular}
}
\end{table}

\section{Conclusion and Future Work}
In this work, we present a pilot study on the streaming decoder-only ASR with discrete speech units. 
We explore two approaches to achieving streaming decoder-only ASR: Text Token Insertion (TTI) and Boundary Token Insertion (BTI). Experimental results on AISHELL-1 and -2 show that the BTI method yields significantly better performance and competitive CER with the non-streaming decoder-only model. With the initialization of pretrained LLM, the performance of our proposed streaming decoder-only model can be further improved. 
As a pilot study, there remains considerable work to be explored in the follow-up.  We will conduct experiments on more languages, larger datasets, and large-scale models. Note that in this study, we use HuBERT as our speech tokenizer, we will also compare more speech tokenizers in the future.


\bibliographystyle{IEEEtran}
\bibliography{mybib}

\begin{thebibliography}{10}
\providecommand{\url}[1]{#1}
\csname url@samestyle\endcsname
\providecommand{\newblock}{\relax}
\providecommand{\bibinfo}[2]{#2}
\providecommand{\BIBentrySTDinterwordspacing}{\spaceskip=0pt\relax}
\providecommand{\BIBentryALTinterwordstretchfactor}{4}
\providecommand{\BIBentryALTinterwordspacing}{\spaceskip=\fontdimen2\font plus
\BIBentryALTinterwordstretchfactor\fontdimen3\font minus \fontdimen4\font\relax}
\providecommand{\BIBforeignlanguage}[2]{{%
\expandafter\ifx\csname l@#1\endcsname\relax
\typeout{** WARNING: IEEEtran.bst: No hyphenation pattern has been}%
\typeout{** loaded for the language `#1'. Using the pattern for}%
\typeout{** the default language instead.}%
\else
\language=\csname l@#1\endcsname
\fi
#2}}
\providecommand{\BIBdecl}{\relax}
\BIBdecl

\bibitem{gpt}
OpenAI, ``{GPT-4} technical report,'' \emph{CoRR}, vol. abs/2303.08774, 2023.

\bibitem{llama}
H.~Touvron, T.~Lavril, G.~Izacard, X.~Martinet, M.~Lachaux, T.~Lacroix, B.~Rozi{\`{e}}re, N.~Goyal, E.~Hambro, F.~Azhar, A.~Rodriguez, A.~Joulin, E.~Grave, and G.~Lample, ``Llama: Open and efficient foundation language models,'' \emph{CoRR}, vol. abs/2302.13971, 2023.

\bibitem{glm}
Z.~Du, Y.~Qian, X.~Liu, M.~Ding, J.~Qiu, Z.~Yang, and J.~Tang, ``{GLM:} general language model pretraining with autoregressive blank infilling,'' in \emph{the 60th Annual Meeting of the Association for ComputationalLinguistics,{ACL} 2022}.\hskip 1em plus 0.5em minus 0.4em\relax Association for Computational Linguistics, 2022, pp. 320--335.

\bibitem{qwen}
J.~Bai, S.~Bai, Y.~Chu, Z.~Cui, K.~Dang, X.~Deng, Y.~Fan, W.~Ge, Y.~Han, F.~Huang, B.~Hui, L.~Ji, M.~Li, J.~Lin, R.~Lin, D.~Liu, G.~Liu, C.~Lu, K.~Lu, J.~Ma, R.~Men, X.~Ren, X.~Ren, C.~Tan, S.~Tan, J.~Tu, P.~Wang, and et~al., ``Qwen technical report,'' \emph{CoRR}, vol. abs/2309.16609, 2023.

\bibitem{videogpt}
W.~Yan, Y.~Zhang, P.~Abbeel, and A.~Srinivas, ``Videogpt: Video generation using {VQ-VAE} and transformers,'' \emph{CoRR}, vol. abs/2104.10157, 2021.

\bibitem{cv1}
J.~Y. Koh, R.~Salakhutdinov, and D.~Fried, ``Grounding language models to images for multimodal inputs and outputs,'' in \emph{International Conference on Machine Learning ,{ICML} 2023}.\hskip 1em plus 0.5em minus 0.4em\relax {PMLR}, 2023, pp. 17\,283--17\,300.

\bibitem{cv2}
L.~Yu, Y.~Cheng, Z.~Wang, V.~Kumar, W.~Macherey, Y.~Huang, D.~A. Ross, I.~Essa, Y.~Bisk, M.~Yang, K.~P. Murphy, A.~G. Hauptmann, and L.~Jiang, ``{SPAE:} semantic pyramid autoencoder for multimodal generation with frozen llms,'' in \emph{Neural Information Processing Systems,NeurIP}, 2023.

\bibitem{viola}
T.~Wang, L.~Zhou, Z.~Zhang, Y.~Wu, S.~Liu, Y.~Gaur, Z.~Chen, J.~Li, and F.~Wei, ``Viola: Unified codec language models for speech recognition, synthesis, and translation,'' \emph{CoRR}, vol. abs/2305.16107, 2023.

\bibitem{speechgpt}
D.~Zhang, S.~Li, X.~Zhang, J.~Zhan, P.~Wang, Y.~Zhou, and X.~Qiu, ``Speechgpt: Empowering large language models with intrinsic cross-modal conversational abilities,'' in \emph{{EMNLP} 2023}.\hskip 1em plus 0.5em minus 0.4em\relax ACL, 2023, pp. 15\,757--15\,773.

\bibitem{audiopalm}
P.~K. Rubenstein, C.~Asawaroengchai, D.~D. Nguyen, A.~Bapna, Z.~Borsos, F.~de~Chaumont~Quitry, P.~Chen, D.~E. Badawy, W.~Han, E.~Kharitonov, H.~Muckenhirn, D.~Padfield, and et~al., ``Audiopalm: {A} large language model that can speak and listen,'' \emph{CoRR}, vol. abs/2306.12925, 2023.

\bibitem{laura}
J.~Wang, Z.~Du, Q.~Chen, Y.~Chu, Z.~Gao, Z.~Li, K.~Hu, X.~Zhou, J.~Xu, Z.~Ma, W.~Wang, S.~Zheng, C.~Zhou, Z.~Yan, and S.~Zhang, ``Lauragpt: Listen, attend, understand, and regenerate audio with {GPT},'' \emph{CoRR}, vol. abs/2310.04673, 2023.

\bibitem{DBLP:journals/corr/abs-2401-10446}
\BIBentryALTinterwordspacing
Y.~Hu, C.~Chen, C.~H. Yang, R.~Li, C.~Zhang, P.~Chen, and E.~S. Chng, ``Large language models are efficient learners of noise-robust speech recognition,'' \emph{CoRR}, vol. abs/2401.10446, 2024. [Online]. Available: \url{https://doi.org/10.48550/arXiv.2401.10446}
\BIBentrySTDinterwordspacing

\bibitem{DBLP:journals/corr/abs-2402-05457}
\BIBentryALTinterwordspacing
C.~Chen, R.~Li, Y.~Hu, S.~M. Siniscalchi, P.~Chen, E.~S. Chng, and C.~H. Yang, ``It's never too late: Fusing acoustic information into large language models for automatic speech recognition,'' \emph{CoRR}, vol. abs/2402.05457, 2024. [Online]. Available: \url{https://doi.org/10.48550/arXiv.2402.05457}
\BIBentrySTDinterwordspacing

\bibitem{transformer}
A.~Vaswani, N.~Shazeer, N.~Parmar, J.~Uszkoreit, L.~Jones, A.~N. Gomez, L.~Kaiser, and I.~Polosukhin, ``Attention is all you need,'' in \emph{Neural Information Processing Systems (NeurIPS)}, 2017, pp. 5998--6008.

\bibitem{qwenaudio}
Y.~Chu, J.~Xu, X.~Zhou, Q.~Yang, S.~Zhang, Z.~Yan, C.~Zhou, and J.~Zhou, ``Qwen-audio: Advancing universal audio understanding via unified large-scale audio-language models,'' \emph{CoRR}, vol. abs/2311.07919, 2023.

\bibitem{salm}
C.~Tang, W.~Yu, G.~Sun, X.~Chen, T.~Tan, W.~Li, L.~Lu, Z.~Ma, and C.~Zhang, ``{SALMONN:} towards generic hearing abilities for large language models,'' \emph{CoRR}, vol. abs/2310.13289, 2023.

\bibitem{wu}
J.~Wu, Y.~Gaur, Z.~Chen, L.~Zhou, Y.~Zhu, T.~Wang, J.~Li, S.~Liu, B.~Ren, L.~Liu, and Y.~Wu, ``On decoder-only architecture for speech-to-text and large language model integration,'' in \emph{{IEEE} Automatic Speech Recognition and Understanding Workshop, {ASRU} 2023}.\hskip 1em plus 0.5em minus 0.4em\relax {IEEE}, 2023, pp. 1--8.

\bibitem{encodec}
A.~D{\'{e}}fossez, J.~Copet, G.~Synnaeve, and Y.~Adi, ``High fidelity neural audio compression,'' \emph{CoRR}, vol. abs/2210.13438, 2022.

\bibitem{hubert}
W.~Hsu, B.~Bolte, Y.~H. Tsai, and et~al., ``Hubert: Self-supervised speech representation learning by masked prediction of hidden units,'' \emph{{IEEE} {ACM} Trans. Audio Speech Lang. Process.}, vol.~29, pp. 3451--3460, 2021.

\bibitem{palm}
R.~Anil, A.~M. Dai, O.~Firat, M.~Johnson, D.~Lepikhin, A.~Passos, S.~Shakeri, E.~Taropa, P.~Bailey, Z.~Chen, E.~Chu, J.~H. Clark, L.~E. Shafey, Y.~Huang, K.~Meier{-}Hellstern, G.~Mishra, E.~Moreira, M.~Omernick, K.~Robinson, S.~Ruder, Y.~Tay, K.~Xiao, Y.~Cheng, C.~Cherry, L.~Gonzalez, and et~al., ``Palm 2 technical report,'' \emph{CoRR}, vol. abs/2305.10403, 2023.

\bibitem{usm}
Y.~Zhang, W.~Han, J.~Qin, Y.~Wang, A.~Bapna, Z.~Chen, N.~Chen, B.~Li, V.~Axelrod, G.~Wang, Z.~Meng, K.~Hu, A.~Rosenberg, R.~Prabhavalkar, D.~S. Park, P.~Haghani, J.~Riesa, G.~Perng, H.~Soltau, T.~Strohman, F.~Beaufays, Y.~Wu, and et~al., ``Google {USM:} scaling automatic speech recognition beyond 100 languages,'' \emph{CoRR}, vol. abs/2303.01037, 2023.

\bibitem{semi}
S.~Arora, G.~Saon, S.~Watanabe, and B.~Kingsbury, ``Semi-autoregressive streaming {ASR} with label context,'' in \emph{{IEEE} International Conference on Acoustics, Speech and Signal Processing, {ICASSP} 2024}.\hskip 1em plus 0.5em minus 0.4em\relax {IEEE}, 2024.

\bibitem{stream2}
Q.~Li, B.~Li, D.~Hwang, T.~N. Sainath, and P.~M. Mengibar, ``Modular domain adaptation for conformer-based streaming {ASR},'' \emph{CoRR}, vol. abs/2305.13408, 2023.

\bibitem{lossmask}
Q.~Chen, W.~Wang, Q.~Zhang, S.~Zheng, S.~Zhang, C.~Deng, Y.~Ma, H.~Yu, J.~Liu, and C.~Zhang, ``Loss masking is not needed in decoder-only transformer for discrete-token based {ASR},'' in \emph{{IEEE} International Conference on Acoustics, Speech and Signal Processing, {ICASSP} 2024}.\hskip 1em plus 0.5em minus 0.4em\relax {IEEE}, 2024.

\bibitem{gmm}
L.~R. Rabiner, ``A tutorial on hidden markov models and selected applications in speech recognition,'' \emph{Proc. {IEEE}}, vol.~77, no.~2, pp. 257--286, 1989.

\bibitem{aishell1}
H.~Bu, J.~Du, X.~Na, B.~Wu, and H.~Zheng, ``{AISHELL-1:} an open-source mandarin speech corpus and a speech recognition baseline,'' in \emph{{O-COCOSDA} 2017}.\hskip 1em plus 0.5em minus 0.4em\relax {IEEE}, 2017, pp. 1--5.

\bibitem{aishell2}
J.~Du, X.~Na, X.~Liu, and H.~Bu, ``{AISHELL-2:} transforming mandarin {ASR} research into industrial scale,'' \emph{CoRR}, vol. abs/1808.10583, 2018.

\bibitem{ls}
C.~Szegedy, V.~Vanhoucke, S.~Ioffe, J.~Shlens, and Z.~Wojna, ``Rethinking the inception architecture for computer vision,'' in \emph{{IEEE} Conference on Computer Vision and Pattern Recognition, {CVPR} 2016}.\hskip 1em plus 0.5em minus 0.4em\relax {IEEE} Computer Society, 2016, pp. 2818--2826.

\bibitem{wenet}
B.~Zhang, D.~Wu, Z.~Peng, X.~Song, Z.~Yao, H.~Lv, L.~Xie, C.~Yang, F.~Pan, and J.~Niu, ``Wenet 2.0: More productive end-to-end speech recognition toolkit,'' in \emph{Interspeech 2022}.\hskip 1em plus 0.5em minus 0.4em\relax {ISCA}, 2022, pp. 1661--1665.

\bibitem{CCA}
A.~Pasad, B.~Shi, and K.~Livescu, ``Comparative layer-wise analysis of self-supervised speech models,'' in \emph{{IEEE} International Conference on Acoustics, Speech and Signal Processing, {ICASSP} 2023}.\hskip 1em plus 0.5em minus 0.4em\relax {IEEE}, 2023, pp. 1--5.

\bibitem{chang}
X.~Chang, B.~Yan, K.~Choi, J.~Jung, Y.~Lu, S.~Maiti, R.~S. Sharma, J.~Shi, J.~Tian, S.~Watanabe, Y.~Fujita, T.~Maekaku, P.~Guo, Y.~Cheng, P.~Denisov, K.~Saijo, and H.~Wang, ``Exploring speech recognition, translation, and understanding with discrete speech units: {A} comparative study,'' in \emph{{IEEE} International Conference on Acoustics, Speech and Signal Processing,{ICASSP} 2024}.\hskip 1em plus 0.5em minus 0.4em\relax {IEEE}, pp. 11\,481--11\,485.

\bibitem{ebran}
K.~Kim, F.~Wu, Y.~Peng, J.~Pan, P.~Sridhar, K.~J. Han, and S.~Watanabe, ``E-branchformer: Branchformer with enhanced merging for speech recognition,'' in \emph{{IEEE} Spoken Language Technology Workshop,{SLT} 2022}.\hskip 1em plus 0.5em minus 0.4em\relax {IEEE}, 2022, pp. 84--91.

\end{thebibliography}

\end{document}